\documentclass[letter]{aa} % for the letters 
\usepackage{graphicx}
\usepackage{txfonts}
\usepackage{hyperref}
\hypersetup{colorlinks=true,linkcolor=blue,citecolor=blue,filecolor=blue,urlcolor=blue}
\newcommand{\ha}{H$\alpha$~}
\newcommand{\bg}{Br$\gamma$~}
\newcommand{\hei}{\mbox{\ion{He}{i} $\lambda$10830~}}
\newcommand{\heic}{\ion{He}{i} $\lambda$10830}
\newcommand{\pg}{Pa$\gamma$~}
\newcommand{\pgc}{Pa$\gamma$}
\begin{document}

   \title{An infrared FWHM-$K_2$ correlation to uncover highly reddened quiescent black holes}
   \titlerunning{Infrared FWHM-$K_2$ correlation}
   \authorrunning{V.~A.~C\'uneo et al.}

   \author{V.~A.~C\'uneo\inst{1,2},
          J.~Casares\inst{1,2},
          M.~Armas~Padilla\inst{1,2},
          J.~S\'anchez-Sierras\inst{1,2},
          J.~M.~Corral-Santana\inst{3,4},
          T.~J.~Maccarone\inst{5},
          D.~Mata~S\'anchez\inst{1,2},
          T.~Mu\~noz-Darias\inst{1,2},
          M.~A.~P.~Torres\inst{1,2},
          \and
          F.~Vincentelli\inst{1,2}
          }

   \institute{Instituto de Astrofísica de Canarias (IAC), Vía Láctea s/n, E-38205 La Laguna, Tenerife, Spain\\
              \email{virginiacuneo@gmail.com}
         \and
             Departamento de Astrofísica, Universidad de La Laguna, Av. Astrofísico Francisco Sánchez s/n, E-38206 La Laguna, Tenerife, Spain
         \and
             European Southern Observatory (ESO), Alonso de C\'ordova 3107, Vitacura, Casilla 19, Santiago, Chile
         \and 
             Pontificia Universidad Cat\'olica de Chile, Vicu\~na-Mackenna 4860, Macul, Santiago, Chile
         \and
             Department of Physics \& Astronomy, Texas Tech University, Box 41051, Lubbock, TX, 79409-1051, USA
             }

   \date{Received 2 October 2023 / Accepted 26 October 2023}
   %\date{Received XXX; accepted YYY}

% \abstract{}{}{}{}{} 
% 5 {} token are mandatory
 
  \abstract
   {Among the sample of Galactic transient X-ray binaries (SXTs) discovered to date, about 70 have been proposed as likely candidates  to host a black hole. Yet, only 19 have been dynamically confirmed. Such a reliable confirmation requires phase-resolved spectroscopy of their companion stars, which is generally feasible when the system is in a quiescent state. However, since most of the SXT population lies in the galactic plane, which is strongly affected by interstellar extinction, their optical brightness during quiescence usually falls beyond the capabilities of the current instrumentation \mbox{($R\gtrsim22$)}. To overcome these limitations and thereby increase the number of confirmed Galactic black holes, a correlation between the full-width at half maximum (FWHM) of the \ha line and the semi-amplitude of the donor’s radial velocity curve ($K_2$) was presented in the past. Here, we extend the FWHM-$K_2$ correlation to the near-infrared (NIR), exploiting disc lines such as \heic, \pgc, and Br$\gamma$, in a sample of dynamically confirmed black-hole SXTs. We obtain \mbox{$K_2 = 0.22(3)$ FWHM}, in good agreement with the optical correlation derived using H$\alpha$. The similarity of the two correlations seems to imply that the widths of \ha and the NIR lines are consistent in quiescence. When combined with information on orbital periods, the NIR correlation  allows us to constrain the mass of the compact object of systems in quiescence by using single-epoch spectroscopy. We anticipate that this new correlation will give access to highly reddened black hole SXTs, which cannot be otherwise studied at optical wavelengths.}
  

   \maketitle
%
%________________________________________________________________
\section{Introduction}
   
Black holes (BHs) have long been subject of intense study due to their display of extreme physics, such as accretion and outflows, which is common to sources at all scales, from X-ray binaries to active galactic nuclei \citep[e.g.][]{Fabian2012,Fender2016}. In particular, stellar-mass BHs provide the ideal laboratories to study the phenomenon of accretion on timescales suitable for human investigation. In addition, BHs with known masses play a key role in testing and understanding the supernova explosion mechanism and the formation of compact objects \citep[e.g.][]{Belczynski2012,Casares2017}.

Four techniques are currently used to find stellar-mass BHs, with each offering some advantages and some limitations relative to the others. Gravitational waves have now detected large samples of BHs \citep[e.g.][]{Abbott2023}, but are strongly biased toward the more massive objects. Orbital measurements of the companion stars in detached BH binaries \citep[e.g.][]{Geisers2018,El-Badry2023} and microlensing detections \citep[e.g.][]{Lam2022,Sahu2022} enable the detection of BHs without strong interactions with their environment, but offer no probes of the BH spins. Transient X-ray binaries \mbox{(so} called soft X-ray transients, \mbox{SXTs)} offer the opportuniy to use X-rays to probe spins \citep[e.g.][]{Reynolds2021}, but they typically form via common envelope evolution and, hence, may not be representative of the global mass distribution. Generally, SXTs are discovered when they enter a usually short-lived but bright outburst phase (see \citealt{Corral-Santana2016} for a review; see also \citealt{McClintock2006}). They host either a neutron star (NS) or a BH that accretes matter from a low-mass \mbox{($\leq1 ~M_{_\odot}$)} donor star through an accretion disc.

\begin{table*}
    \centering
        \caption{Sample of SXTs used in this work.}
        \label{obs}
        %\resizebox{\textwidth}{!}{
        \begin{tabular}{lccccccc} 
                \hline
                \rule{0pt}{2.5ex}\bf{X-ray transient} & Type & Spectra & Observing period & Resolution & Spectral & Telescope/ & References\\
                 &  &  &  & (km s$^{-1}$) & band & instrument & \\
        \\[-2ex]
                \hline
                \rule{0pt}{2.5ex}Nova Mus 1991 & BH & 17 $\times$ 1200s & 13/04/13 - 10/05/13 & 70 & J, H, K & VLT/X-shooter & 1 \\
                \rule{0pt}{2ex}A0620-00 & BH & 12 $\times$ 240s & 17/02/05 & 150 & K & Keck II/NIRSPEC & 2 \\
        \rule{0pt}{2ex}XTE J1118$+$480 & BH & 48 $\times$ 310s & 02/04/11 and 12/04/11 & 176 & J, H, K & Gemini/GNIRS & 3 \\
                \rule{0pt}{2ex}GX 339-4 & BH & 16 $\times$ 275s & 22/05/16 - 07/09/16 & 55 & J, H & VLT/X-shooter & 4 \\
                \rule{0pt}{2ex}GRS 1915$+$105 & BH & 27 $\times$ 2400s & 09/06/10 - 03/09/11 & 37 & J, H, K & VLT/X-shooter & 5 \\
                \rule{0pt}{2ex}Aql X-1 & NS & 24 $\times$ 900s & 20/05/10 - 29/09/11 & 75 & K & VLT/SINFONI & 6 \\
                \hline
        \end{tabular}%}
    \tablebib{(1) \citet{GonzalezHernandez2017}; (2) \citet{Harrison2007}; (3) \citet{Khargharia2013}; (4) \citet{Heida2017}; (5) \citet{Steeghs2013}; (6) \citet{MataSanchez2017}.}
\end{table*}    
   
To date, $\sim$70 BH candidates have been detected in SXTs \citep[e.g.][]{Corral-Santana2016,Tetarenko2016}. However, an empirical extrapolation of that sample implies that about 2000 BH SXTs are expected to exist in the Galaxy \citep{Romani1998,Corral-Santana2016}, while some population-synthesis models predict that the number ought to be $\geq10^4$ \citep{Kiel2006,Yungelson2006}. In addition to the low detection statistics, determining the BH nature of the compact object in these systems represents a problem in itself. We need to estimate the mass of the compact component in order to confirm its nature, which requires phase-resolved spectroscopy to perform optical dynamical studies of the low-mass companion star \citep[e.g.][]{Casares1992,Torres2019}. Given that during outburst the accretion disc is so bright that it conceals entirely the companion star, these studies are done in quiescence, when the optical brightness of the binary is dominated by the donor, with the addition of broad emission lines that evidence the presence of the accretion disc \citep[e.g.][]{Charles2006}. On the one hand, this represents an advantage, since SXTs spend most of the time in the quiescent state. On the other hand, the companion is a low-mass, and therefore an intrinsically faint, star whose apparent brightness usually falls beyond the observing threshold of the largest available telescopes, namely, $R\gtrsim22.  $  We refer to \citet{Corral-Santana2011} and \citet{Yanez-Rizo2022} for examples of dynamical studies of the companion star at the limit of the instrumental capabilities. We are therefore biased towards detecting BHs only in the brightest and closest systems. As a result, from the sample of BH candidates, only 19 have been dynamically confirmed (\citealt{Casares2014,Corral-Santana2016}; see online version of BlackCAT\footnote{BlackCAT: \url{https://www.astro.puc.cl/BlackCAT/}}). 

Based on the fact that the emission lines from the quiescent disc, particularly H$\alpha$, are much stronger (with larger equivalent widths) than the absorption lines from the companion star, \citet{Casares2015} proposed a novel technique to determine the mass of a BH. This approach can reach systems \mbox{$\sim$2.5 mag} fainter than with the classical dynamical studies. The method, which relies on a scaling correlation between the full width at half-maximum (FWHM) of \ha and the projected velocity semi-amplitude of the companion star ($K_2$), allows us to uncover new \mbox{BH SXTs} by solely resolving \ha and knowing the orbital period \citep[\textit{P}\textsubscript{orb};][]{MataSanchez2015,Casares2018}. 

SXTs with quiescent \textit{R} magnitudes fainter than \mbox{$\sim$22-24} can either be intrinsically faint or extincted by interstellar material. Most of the detected BH SXT candidates are located along the galactic plane, where interstellar extinction is usually high. The infrared spectral region is significantly less affected by the extinction \citep[e.g.][]{Wang2019} and it is therefore appropriate for investigating new techniques to uncover BHs \citep[e.g.][]{Steeghs2013}. In this work, we extend the method from \citet{Casares2015} to the near-infrared (NIR) regime. To this aim, we investigate the typical emission lines present in the NIR spectra of known SXTs. The NIR FWHM-$K_2$ correlation allows us to measure masses of the compact objects in optically faint SXTs, based on single-epoch spectroscopy datasets.

%__________________________________________________________________
\section{The sample}

We compiled a spectroscopy sample of five dynamically confirmed BH SXTs from the BlackCAT catalogue and a NS SXT, exhibiting \heic, \pg and/or \bg in emission during the quiescent state. \mbox{Table \ref{obs}} contains a summarised observing log for each system, including references to the literature with the original publication for further details. We downloaded and reduced the data for \mbox{A0620-00} \citep[first published by][]{Harrison2007} using the \textsc{wmkonspec} package and usual \textsc{iraf}\footnote{IRAF is distributed by the National Optical Astronomy Observatory, which is operated by the Association of Universities for Research in Astronomy, Inc. under contract to the National Science Foundation.} tasks. We used the \textsc{gemini iraf} package to process the \mbox{XTE J1118$+$480} spectra, first presented in \citet{Khargharia2013}. In addition, we downloaded the \mbox{Nova Mus 1991}, \mbox{GX 339-4}, and \mbox{GRS 1915$+$105} data published in \citet{GonzalezHernandez2017}, \citet{Heida2017}, and \citet{Steeghs2013}, respectively, and processed them using the ESO X-shooter pipeline v.3.5.0. In the case of \mbox{Aql X-1}, we used the spectra published and reduced in \citet{MataSanchez2017}.

We acknowledge that since its discovery outburst in 1992, \mbox{GRS 1915$+$105} was never in a fully quiescent state. Its X-ray flux was highly variable until 2018, when it decayed to a low-flux plateau \citep[e.g.][]{Motta2021}. Nonetheless, the large size of the accretion disc (\mbox{\textit{P}\textsubscript{orb} $=33.85 \pm 0.16$ days}) and the fact that the donor star is not affected by irradiation \citep{Steeghs2013}, imply that irradiation in the outer disc (where the emission lines that we analyse are formed; see Section \ref{ana}) is modest. In addition, most of the data used in this study were obtained in periods of faint X-ray emission. Therefore, we expect the emission lines included in our analysis to behave like in a quiescent disc. 

\hei and \pg are in relatively clean spectral regions. However, \bg is in a spectral region where the atmosphere contribution might be significant, which requires the application of techniques for correction from telluric features. The \mbox{GRS 1915$+$105} spectra were corrected from telluric absorptions of H\textsubscript{2}O and CH\textsubscript{4} molecules using \textsc{molecfit} v.3.0.3. The use of \textsc{iraf} tasks together with spectra from a telluric star was undertaken for the correction of A0620-00 spectra, while custom software under \textsc{python} was used for the case of \mbox{Aql X-1}.

%______________________________________________________________

\begin{figure*}
 \includegraphics[width=\textwidth]{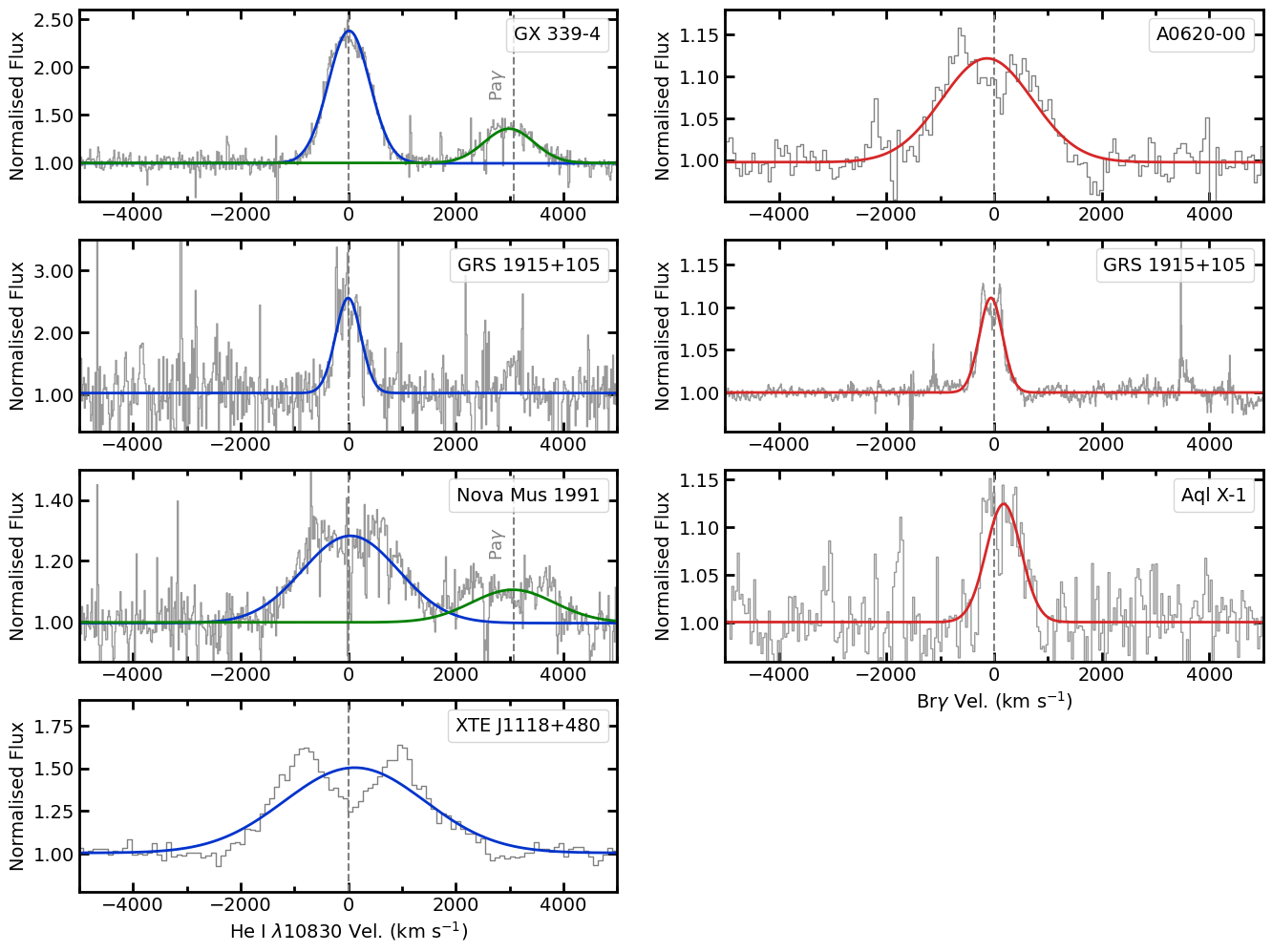}
 \caption{Gaussian fits to the NIR emission line profiles. The left column shows the fits for the \hei (blue) and \pg (green; when the S/N allowed to fit the line) emission lines, while the fits to \bg (red) are shown in the right column.}
 \label{fits}
\end{figure*}

%______________________________________________________________
\section{The FWHM-$K_2$ correlation}
\label{ana}

We focused our analysis on the strongest NIR transitions present in the data (\heic, \pg and/or Br$\gamma$) using \textsc{molly} and custom software under \mbox{\textsc{python} 3.7}. We carefully normalised each emission line by fitting the adjacent continuum with a low order polynomial and we combined the spectra to obtain one single spectrum per line and source with a higher signal-to-noise ratio (S/N). The FWHM values were obtained from fitting each emission line with a model consisting of a Gaussian profile plus a constant, over a spectral region of \mbox{$\pm$10000 km s$^{-1}$} around the line rest wavelength (with the exception of \mbox{GRS 1915$+$105}, for which we used a spectral region of \mbox{$\pm$4000 km s$^{-1}$} to avoid contamination by nearby instrumental features). We used orthogonal distance regression in the ordinary least squares mode, implementing the \textsc{python} package \textsc{scipy.odr}, to perform these fits. We notice that some of the lines present the double-peaked profile caused by the rotation of the disc \citep{Smak1969}. However, following \citet{Casares2015}, we decided to fit a Gaussian to all the data, which is a more robust model than a double-Gaussian. While the double-Gaussian might fail to fit data of limited spectral resolution and poor S/N, \citet{Casares2015} found that the FWHM values obtained from a single-Gaussian are within a \mbox{10$\%$} of those obtained with more complex models. In the case of \mbox{\heic}, the neighbouring \pg emission line was masked when present -- and vice versa. 

Spectral variations from one epoch to another may be present and can be caused by aperiodic flares and orbital changes, such as disc asymmetries \citep[][]{Hynes2002,Casares2015}. However, in most cases we are limited by the S/N, which can reflect the observing conditions of the night. In order to estimate the impact of additional systematics related to the intrinsic variability of the FWHM we used the spectra of \mbox{GX 339-4}, which exhibit the highest S/N data in our sample. A Gaussian fit allowed us to estimate the FWHM value of the \mbox{\hei} line, along with its corresponding standard deviation, for each of the 16 spectra individually. We then used both the error propagation and the standard deviation to obtain a more conservative error \mbox{(of $\sim$10$\%$)} for the combined data of \mbox{GX 339-4}. Then, for the remaining sources of the sample we estimated the error of the FWHM as the quadratic sum of the statistical error from the fit and the \mbox{10$\%$} of the FWHM. To obtain the intrinsic FWHMs, we subtracted  the instrumental resolution quadratically. \mbox{Figure \ref{fits}} displays the fits to each line included in our sample. In \mbox{Table \ref{sample},} we list the FWHM values we measured, as well as the residual variance ($\sigma^2_r$), which parameterises the quality of the Gaussian fits by quantifying deviations between the data and the best-fitting model, as well the dynamical $K_2$ values with their corresponding references. We note that the reported value of $\sigma^2_r$ for the fit of the \hei line in \mbox{GX 339-4} in \mbox{Table \ref{sample}} is an average of the 16 $\sigma^2_r$ values from the individual fits.

\begin{table*}
        \centering
        \caption{SXT parameters.}
        \label{sample}
        \resizebox{\textwidth}{!}{
        \begin{tabular}{lccccccccc} 
                \hline
                \rule{0pt}{2.5ex}\bf{X-ray transient} & Emission & FWHM & $\sigma^2_r$ & $K_2$ & $K_2$ from corr. & \textit{P}\textsubscript{orb} & $f(M)$ & PMF & Ref. \\
                 & line & (km s$^{-1}$) &  & (km s$^{-1}$) & (km s$^{-1}$) & (days) & ($M_{\odot}$) & ($M_{\odot}$) & \\
        \\[-2ex]
                \hline
                \rule{0pt}{2.5ex}Nova Mus 1991 & \hei & 2097 $\pm$ 213 & 26.07 & 406.8 $\pm$ 2.7 & 461 $\pm$ 78 & 0.43260249(9) & 3.02 $\pm$ 0.06 & 4.4 $\pm$ 2.2 & (1) \\
        \rule{0pt}{2ex} & \pg & 1811 $\pm$ 197 & 26.06 &  & 398 $\pm$ 69 &  &  & 2.8 $\pm$ 1.5 &  \\
                \rule{0pt}{2ex}A0620-00 & \bg & 1974 $\pm$ 204 & 4.31 & 437.1 $\pm$ 2.0 & 434 $\pm$ 74 & 0.32301405(1) & 2.79 $\pm$ 0.04 & 2.7 $\pm$ 1.4 & (2) \\
                \rule{0pt}{2ex}XTE J1118$+$480 & \hei & 3078 $\pm$ 310 & 20.06 & 708.8 $\pm$ 1.4 & 677 $\pm$ 115 & 0.1699338(5) & 6.27 $\pm$ 0.04 & 5.5 $\pm$ 2.8 & (3) \\
                \rule{0pt}{2ex}GX 339-4 & \hei & 900 $\pm$ 77 & 4.18 & 219.0 $\pm$ 3.0 & 198 $\pm$ 32 & 1.7587(5) & 1.91 $\pm$ 0.08 & 1.4 $\pm$ 0.7 & (4) \\
        \rule{0pt}{2ex} & \pg & 1088 $\pm$ 111 & 5.76 &  & 239 $\pm$ 41 &  &  & 2.5 $\pm$ 1.3 &  \\
                \rule{0pt}{2ex}GRS 1915$+$105 & \hei & 551 $\pm$ 59 & 6.72 & 126.0 $\pm$ 1.0 & 121 $\pm$ 21 & 33.85(16) & 7.02 $\pm$ 0.17 & 6.2 $\pm$ 3.2 & (5) \\
        \rule{0pt}{2ex} & \bg & 503 $\pm$ 50 & 44.34 &  & 111 $\pm$ 19 &  &  & 4.8 $\pm$ 2.4 &  \\
                \rule{0pt}{2ex}Aql X-1 & \bg & 762 $\pm$ 86 & 2.45 & 136.0 $\pm$ 4.0 & 168 $\pm$ 30 & 0.7895126(10) & 0.21 $\pm$ 0.02 & 0.4 $\pm$ 0.2 & (6) \\
                \hline
        \end{tabular}}
    \tablebib{(1) \citet{Wu2015}; (2) \citet{GonzalezHernandez2010}; (3) \citet{GonzalezHernandez2012}; (4) \citet{Heida2017}; (5) \citet{Steeghs2013}; (6) \citet{MataSanchez2017}.}
\end{table*}

\begin{figure}
 \includegraphics[width=\hsize]{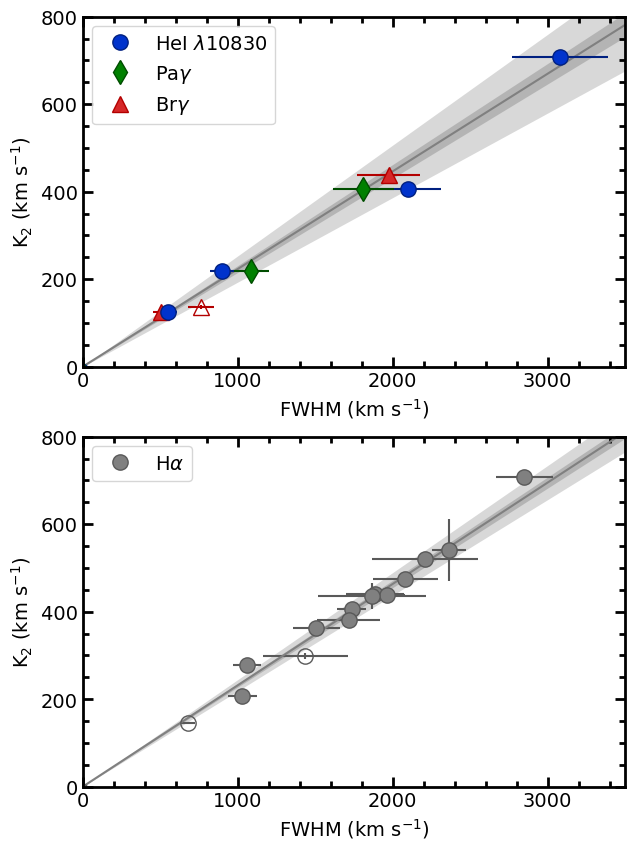}
 \caption{FWHM-$K_2$ correlation for SXTs with the best linear fit, using NIR (top panel) and optical (bottom panel; data come from \citealt{Casares2015}) emission lines. Filled symbols indicate BH binaries, while NS binaries are denoted with open symbols. The shaded regions in both plots correspond to the $1\sigma$ uncertainty of the linear fit (dark grey) and the total uncertainty of the correlation that includes the result from the Monte Carlo simulation in Section \ref{ana} (light grey).}
 \label{corr}
\end{figure}

In the top panel of \mbox{Fig. \ref{corr},} we display the FWHM values versus $K_2$ for the SXTs in our sample. A linear fit, using orthogonal distance regression to account for errors in both FWHM and $K_2$ values, yields the following relation:
\begin{equation}
\label{corr_eq}
    K_2 = 0.22(3) ~\textrm{FWHM},
\end{equation}
where both $K_2$ and FWHM are given in \mbox{km s$^{-1}$}. The residual variance is $\sigma^2_r = 1.04$. Allowing for a constant term does not improve the fit since it is consistent with zero at \mbox{1$\sigma$}. We computed the $K_2$ values from this relation (see \mbox{Table \ref{sample}}) and compared them with the dynamical estimations, finding that the differences follow a Gaussian distribution with a standard deviation of \mbox{26 km s$^{-1}$}. We estimated the error of the coefficient in \mbox{Eq. \ref{corr_eq}} as the quadratic sum of the statistical error from the fit and the result from a Monte Carlo simulation of \mbox{10$^4$} events, imposing the condition that the difference between the model and the real $K_2$ values follow a Gaussian distribution with \mbox{$\sigma = 26$ km s$^{-1}$}. It is worth mentioning that although \mbox{GRS 1915$+$105} was not in full quiescence, and its FWHM values define the bottom end of the correlation, the best-fit parameters remain within the error if we remove those points from the fit.

\citet{Casares2015} derived an analogous relation in the optical ({\mbox{$K_2 = 0.233(13)$ FWHM}}) using the FWHM of the H$\alpha$ emission line. We show this data in the bottom panel of \mbox{Fig. \ref{corr}}. We notice that our NIR correlation \mbox{(Eq. \ref{corr_eq})} is fully compatible with that of H$\alpha$. Furthermore, a single fit to the combined \ha and NIR FWHM values yields best-fit parameters that are fully consistent with both the optical and NIR correlations.

As explained in Section 3 of \citet{Casares2015}, the FWHM-$K_2$ correlation is expected from basic equations. Particularly, for BH SXTs with a typical mass ratio \mbox{$q = 0.1$}, \mbox{$K_2/\textrm{FWHM} = 0.36 \sqrt\alpha$}, where $\alpha$ is the ratio between the disc radius at which the FWHM is determined by gas velocity (R$_W$) and the effective radius of the compact object's Roche lobe (R$_{L1}$). Our correlation entails that the FWHM of the NIR emission lines trace the disc velocity at \mbox{37$\%$ R$_{L1}$}, implying that these lines are formed in the outer regions of the disc (as is the case for H$\alpha$).

%______________________________________________________________
\section{Discussion}

With the aim of uncovering and studying the reddened population of BHs in the Galaxy, in this work, we compiled the NIR spectra of dynamically studied BH SXTs and a system with a NS during quiescence. We found a correlation between the FWHM of \mbox{\heic}, \pgc, and \bg emission lines, and $K_2$ (upper panel in \mbox{Fig. \ref{corr}}). Despite the low statistics due to the limited availability of NIR spectra of SXTs in quiescence, a comparison with the analogous relation found by \citet{Casares2015} at optical wavelengths, using the \ha emission line (bottom panel of \mbox{Fig. \ref{corr}), offers evidence of}  their resemblance, turning the NIR correlation into a reliable tool. Even the correlation of the combined optical and NIR data agrees with the two separate correlations, as mentioned in \mbox{Section \ref{ana}}.

These correlations allow one to obtain $K_2$ by measuring the FWHM in quiescence of either \ha or one of the NIR emission lines used in this study. This method only requires one single spectrum with an emission line, which saves considerable observing time when compared to the phase-resolved spectroscopy needed by the classical method (radial velocities) to estimate $K_2$. We note, however, that a large orbital coverage would be ideal to avoid potential orbital variations. Nonetheless, an error equal to the quadratic sum of \mbox{10\%} of the FWHM and the statistical error (see \mbox{Section \ref{ana}}) can be assumed in the FWHM values when a single spectrum is available.

Most importantly, the FWHM-$K_2$ correlations are fundamental to uncover the hidden (reddened) population of quiescent BHs in the Galaxy. By combining \mbox{Eq. \ref{corr_eq}} with \textit{P}\textsubscript{orb}, we can use Kepler's third law to estimate a preliminary mass function \mbox{$\textrm{PMF}=1.1 \times 10^{-9} ~P_{orb} ~\textrm{FWHM}^3$}, where PMF is in units of $M_{\odot}$, \textit{P}\textsubscript{orb} in days, and FWHM in \mbox{km s$^{-1}$} (see \citealt{Casares2015} for the optical analogue). The PMF gives a lower limit to the mass of the compact component in the binary system from single epoch spectroscopy. For \mbox{PMF $\gtrsim$ 3 M$_{\odot}$}, the assumed approximate mass limit to segregate BHs from NSs \citep[e.g.][]{Kalogera1996b,Rezzolla2018}, we can positively confirm the BH nature of the compact object. \mbox{Table \ref{sample}} includes \textit{P}\textsubscript{orb}, as well as our PMF estimations and, for comparison, the dynamical mass functions ($f(M)$) with the corresponding references. As expected, the latter two are fully consistent within the uncertainties.

The similarity of the optical and NIR correlations implies that the widths of \ha and the NIR lines, during quiescence, are consistent. This result was expected, since both \ha and the NIR emission lines are tracing the velocity from similar outer regions of the disc (\mbox{$\sim$40$\%$ R$_{L1}$}; see \mbox{Section \ref{ana}}). The profiles of spectral disc emission lines from SXTs during outburst have shown both similarities (e.g. between \ha and \mbox{\ion{He}{i} $\lambda$10830)} and differences \citep[e.g. between \bg with \mbox{\ion{He}{i} $\lambda$10830};][]{Sanchez-Sierras2020} that might be caused by the presence of outflows. In quiescence, however, given the lack of strong accretion activity, we expect the hydrogen and helium transitions to remain mostly unaltered. In the cases were we could measure the width of more than one emission line, we observe that the values are consistent within errors. In order to analyse the correlation for these transitions separately, additional simultaneous data are required.

Using a sample of cataclysmic variables, \citet{Casares2015} showed that the slope of the FWHM-$K_2$ correlation decreases with increasing mass ratio \textit{q}. We note that one of our systems, \mbox{Aql X-1}, is a NS SXT with \mbox{$q = 0.41 \pm 0.08$} \citep{MataSanchez2017}, while the others are BH SXTs with typical mass ratios \mbox{$q \sim 0.1$}. This may explain why \mbox{Aql X-1} lies slightly under the correlation in \mbox{Fig. \ref{corr}}. However, a single data point is not enough to make strong assumptions, and more observations of NS SXTs are needed for confirmation.

Following with exploiting relations to derive fundamental parameters from faint SXTs, future work should employ NIR emission lines to explore the correlations previously established using H$\alpha$. In particular, \citet{Casares2016} reported a correlation between the mass ratio (\textit{q}) and the ratio of the double-peaked separation to the FWHM of H$\alpha$. Subsequently, a correlation between the binary inclination and the depth of the trough from the double-peaked \ha was presented by \citet{Casares2022}.

%______________________________________________________________
\section{Conclusions}
In this work, we compiled quiescent NIR spectra of SXTs \mbox{---mainly BHs---} with known $K_2$ values, and measured the FWHM of \mbox{\heic}, \pg and/or \bg emission lines. We present a NIR FWHM-$K_2$ correlation that allow us to estimate $K_2$ of faint SXTs from single-epoch spectroscopy, saving substantial observing time when compared to the phase-resolved radial velocity method. We find that this NIR correlation is fully consistent with its optical analogue. This result, together with the similarity of the FWHM values of different NIR lines in a given system, suggests that any H or \mbox{\ion{He}{i}} transition could be useful for estimating $K_2$. Most importantly, we are then able to constrain the mass function of the compact object in the system by combining the FWHM-$K_2$ relation with \textit{P}\textsubscript{orb}. This new NIR correlation will not only reveal the nature of highly-reddened SXTs, but will also expand the Galactic BH statistics, allowing for detailed studies of their formation mechanisms.

%__________________________________________________________________
\begin{acknowledgements}
      We thank the anonymous referee for their useful and thoughtful comments, which helped to improve this paper.
      This work is supported by the Spanish Ministry of Science via the Plan de Generacion de conocimiento: PID2020--120323GB--I00 and PID2021-124879NB-I00, and an Europa Excelencia grant (EUR2021-122010). We acknowledge support from the Consejería de Economía, Conocimiento y Empleo del Gobierno de Canarias and the European Regional Development Fund (ERDF) under grant with reference ProID2021010132.\\
    
      We would like to thank D. Steeghs for providing the spectra of GRS 1915$+$105. We are also very grateful to Elizabeth J. Gonzalez for helpful discussions. \\
      
      This research has made use of the Keck Observatory Archive (KOA), which is operated by the W. M. Keck Observatory and the NASA Exoplanet Science Institute (NExScI), under contract with the National Aeronautics and Space Administration. We thank F. A. Cordova, PI of the A0620-00 dataset obtained through KOA.
      This research is based on observations (program ID GN-2011A-Q-13) obtained at the international Gemini Observatory, a program of NSF’s NOIRLab, which is managed by the Association of Universities for Research in Astronomy (AURA) under a cooperative agreement with the National Science Foundation on behalf of the Gemini Observatory partnership: the National Science Foundation (United States), National Research Council (Canada), Agencia Nacional de Investigación y Desarrollo (Chile), Ministerio de Ciencia, Tecnología e Innovación (Argentina), Ministério da Ciência, Tecnologia, Inovações e Comunicações (Brazil), and Korea Astronomy and Space Science Institute (Republic of Korea). 
      This research is also based on observations collected at the European Southern Observatory under ESO programme(s) 091.D-0921(A), 085.D-0497(A) and 085.D-0271(A), and data obtained from the ESO Science Archive Facility under programme(s) 097.D-0915(A) and 297.D-5048(A).\\
      
      \textsc{molly} software (\url{http://deneb.astro.warwick.ac.uk/phsaap/software/molly/html/INDEX.html}) developed by Tom Marsh is gratefully acknowledged. We made use of \texttt{numpy} \citep{Harris2020}, \texttt{astropy} \citep[][]{Robitaille2013,Price-Whelan2018}, \texttt{scipy} \citep{Virtanen2020} and \texttt{matplotlib} \citep{Hunter2007} \textsc{python} packages. We also used \textsc{iraf} \citep{Tody1986} extensively.

\end{acknowledgements}

%-------------------------------------------------------------------
\bibliographystyle{aa} % style aa.bst
\bibliography{references} 

\end{document}